\documentclass[conference]{IEEEtran}
\IEEEoverridecommandlockouts

\usepackage{cite}
\usepackage{amsmath,amssymb,amsfonts}
\usepackage{algorithmic}
\usepackage{graphicx}
\usepackage{textcomp}
\usepackage{xcolor}
\usepackage{url}
\usepackage{booktabs}
\usepackage{breqn}
\def\BibTeX{{\rm B\kern-.05em{\sc i\kern-.025em b}\kern-.08em
    T\kern-.1667em\lower.7ex\hbox{E}\kern-.125emX}}

\usepackage[utf8]{inputenc} 
\usepackage[T1]{fontenc}    
\usepackage[hidelinks]{hyperref} 
\usepackage{url}            
\usepackage{booktabs}       
\usepackage{amsfonts}       
\usepackage{nicefrac}       
\usepackage{microtype}      

\usepackage{amsmath}
\usepackage{amsthm}
\theoremstyle{plain}

\usepackage{subfigure}
\usepackage{graphicx}
\usepackage{tabulary}
\usepackage{colortbl}
\usepackage{multicol}
\usepackage{multirow}
\usepackage{diagbox}
\usepackage{array}
\usepackage{geometry}
\usepackage{longtable}

\begin{document}

\title{FinML-Chain: A Blockchain-Integrated Dataset for Enhanced Financial Machine Learning}

\author{
\IEEEauthorblockN{Jingfeng Chen\textsuperscript{\dag}, Wanlin Deng\textsuperscript{\dag}, Dangxing Chen\textsuperscript{*}, and Luyao Zhang\textsuperscript{*}}
\IEEEauthorblockA{\textit{Duke Kunshan University} \\
Suzhou, China, 215316\\
\thanks{\textsuperscript{\dag}Joint First Authors.}\\
\thanks{\textsuperscript{*}Corresponding Authors: Luyao Zhang (lz183@duke.edu), Data Science Research Center and Social Science Division; Dangxing Chen (dangxing.chen@dukekunshan.edu.cn), Zu Chongzhi Center for Mathematics and Computational Sciences.}\\
\thanks{Address: Duke Kunshan University, Duke Avenue No.8, Kunshan, Suzhou, Jiangsu, China, 215316.}
}
}


\maketitle

\begin{abstract}
Machine learning has become essential for innovation and efficiency in financial markets, offering predictive models and data-driven decision-making. However, challenges such as missing data, lack of transparency, untimely updates, insecurity, and incompatible data sources hinder its effectiveness. These limitations reduce the accuracy and reliability of predictive models. Blockchain technology, with its transparency, immutability, and real-time updates, offers solutions to these challenges. In this paper, we introduce not just a dataset but a new framework for integrating high-frequency on-chain data with low-frequency off-chain data, providing a benchmark for addressing novel research questions in economic mechanism design. This framework enables the generation of modular, extensible datasets tailored to advancing the analysis of economic mechanisms such as the Transaction Fee Mechanism, facilitating multi-modal insights and fairness-driven evaluations. Through time series analysis using four machine learning techniques—linear regression, deep neural networks, XGBoost, and LSTM models—we demonstrate the framework’s capability to produce datasets that drive innovation in financial research and enhance the understanding of blockchain-driven economic systems.

Our contributions are threefold: first, we propose a novel research scenario for the Transaction Fee Mechanism and demonstrate how our framework can address previously unexplored questions in economic mechanism design. Second, we provide a new benchmark for financial machine learning by open-sourcing both the sample dataset generated by our framework and the code for the pipeline, allowing future researchers to append new data and expand the datasets continuously. Third, by ensuring that the framework and its outputs are fully open-source, we promote reproducibility, transparency, and collaboration within the research community. This initiative enables researchers to extend our work, tackle a broader range of economic challenges, and develop innovative financial machine-learning models. By offering this framework, sample datasets, and accompanying resources, we aim to establish a foundation for future advancements in interdisciplinary research at the intersection of machine learning, blockchain, and economics.

\end{abstract}


\section{Introduction}
Financial machine learning has emerged as a pivotal tool for driving innovation and enhancing efficiency in financial markets. By enabling sophisticated predictive models and data-driven decision-making, it offers the potential to significantly improve market outcomes. However, the application of machine learning is fraught with challenges, including missing data, lack of transparency, untimely updates, data insecurity, and the diversity and incompatibility of data sources \cite{hoepner2020significance,8290925,9064510}. These issues collectively impair the accuracy and reliability of predictive models, posing substantial barriers to progress in the field. Blockchain technology presents a novel solution to these challenges. As a distributed database, blockchain offers unique attributes of transparency, immutability, and real-time updates, making it an ideal candidate for addressing the data issues plaguing financial machine learning. Blockchain's decentralized nature ensures that data is transparently recorded and verified by multiple parties, enhancing data reliability and trustworthiness \cite{Andoni2019, bct}. The immutability of blockchain records prevents data tampering and fraud, providing a secure foundation for financial analysis. Real-time updates facilitated by blockchain technology ensure that data is consistently current, enabling timely and accurate model predictions.

In our research, we developed a comprehensive dataset that integrates high-frequency on-chain transaction data with low-frequency off-chain discussion data. The on-chain data ensures traceability, transparency, and security, while the inclusion of off-chain data supports the compatibility and corroboration of different data sources. To assess the potential applicability of this dataset in real-world financial market analysis, we propose the following research questions:
 
\textbf{RQ:} Can this dataset be used to apply different machine learning models for researching innovative financial problems?

We provide all data and code in the following link: \href{https://huggingface.co/datasets/StevenJingfeng/FinML}{https://huggingface.co/datasets/StevenJingfeng/FinML}.

The dataset is 80.4MB for the discord data and 4.92 MB, 13.4MB for two of the on-chain data. Under the property of Blockchain, the data can be continuously updated, and new datasets can be generated at any time. The workflow doesn't specify the size because anyone can extend it.

\subsection{Innovative Mechanism-designed Related Question}

To test the potential of this datatset, we apply it to research how to adavance the latest transaction mechanism implemented in Ethereum. Transaction fee mechanism (TFM) is an essential component of a blockchain protocol to help allocate the limited blockchain computing resources \cite{10.1145/3548606.3559341}. People will consume Ethereum gas and pay for their gas usage if they want to execute transactions in Ethereum. The introduction of EIP-1559 in 2021 restructured user fees into two components: the base fee and the priority fee \cite{10.1145/3548606.3559341}. The base fee reflects the cost of executing the transactions and will be removed from circulation, which is constantly adjusted based on network congestion, aiming to keep block sizes within a target range. The base fee is determined through a Markov process, which involves calculations based on the base fee and gas usage of the previous block. However, due to the calculation formula for the base fee of the next block involving the actual gas consumption of the previous block, EIP-1559 is limited to adjusting its strategy based on events that have already occurred. In other words, the current mechanism of EIP-1559 adjusts the gas price after transactions have taken place. Such a mechanism is passive and cannot proactively predict and strategically adjust the base fee based on upcoming transactions to achieve the mechanism's goal to control the gas used of each block to half of the maximum amount it can be consumed. Zhang~\cite{zhang2023machine} emphasizes the urgent need for AI's fluid adaptability. If we can apply machine learning methods to realize the precise prediction of the gas price for the upcoming transaction and adjust the base fee based on these predictions, we can alter the TFM from ex-post adjustment to proactive adjustment, thus improving the flexibility and efficiency of the transaction fee. Also, the increasing fluid adaptability will contribute to better catering to changes in network traffic and user demands.

The Ethereum EIP-1559 TFM provides an effective scenario for testing our dataset. We aim to test if we can leverage the information within this dataset to achieve reliable future gas demand predictions, which can help facilitate the transition of EIP-1559 from a reactive to a proactive mechanism.

\subsection{Our Approach and Contributions}

\subsubsection{Dataset Validation through Machine Learning Models}  
We employed this dataset to train several machine learning models, including linear regression, deep neural networks (DNN), XGBoost, and Long-Short Term Memory (LSTM) models, to assess the feasibility of predicting gas usage in the next block. This approach validated the dataset’s effectiveness in supporting gas usage prediction.

\subsubsection{Exploration of Multi-Task Capabilities}  
Furthermore, we explored the potential of integrating monotonicity constraints and the BERT model to evaluate the dataset’s capability for handling multiple tasks. These methodologies assess the dataset's compatibility and scalability, ensuring its robustness and applicability in various financial prediction tasks.

\subsubsection{Our Contributions}
\begin{itemize}
    \item \textbf{Introduction of a Novel Framework and Benchmark Dataset}:  
    We developed a new framework that integrates high-frequency on-chain data with low-frequency off-chain data, providing a benchmark for addressing novel research questions in economic mechanism design. This framework generates modular and extensible datasets tailored to financial machine learning, overcoming the challenges of transparency, reliability, and timeliness faced in traditional datasets.

    \item \textbf{Innovative Research Scenario}:  
    We proposed a novel research scenario: leveraging machine learning to optimize blockchain-based transaction fee mechanisms, such as Ethereum's Transaction Fee Mechanism, moving beyond reactive approaches to proactive design. This scenario highlights the practical potential of the framework in advancing mechanism design and other economic applications.

    \item \textbf{Open-Sourcing Data and Pipeline}:  
    To ensure reproducibility and foster collaboration, we openly release not only a sample dataset generated by the framework but also the full pipeline code. This enables researchers to continuously expand the dataset, adapt it to new contexts, and explore a broader range of financial and blockchain-related challenges.
\end{itemize}

Additionally, our framework supports machine learning predictions for demand and supply forecasting, trading mechanism design, and other financial applications. By establishing a benchmark for financial machine learning, it drives innovation and enhances the accuracy and reliability of financial models in the blockchain context.

It is important to note that our contribution is not merely incremental but a qualitative leap. The framework enables the creation of high-frequency, verifiable, and extensible datasets, fundamentally differing from existing benchmarks. Traditional comparisons with existing datasets are thus not applicable in this case.

\begin{figure*}[ht]
\centering
\includegraphics[width=0.9\textwidth]{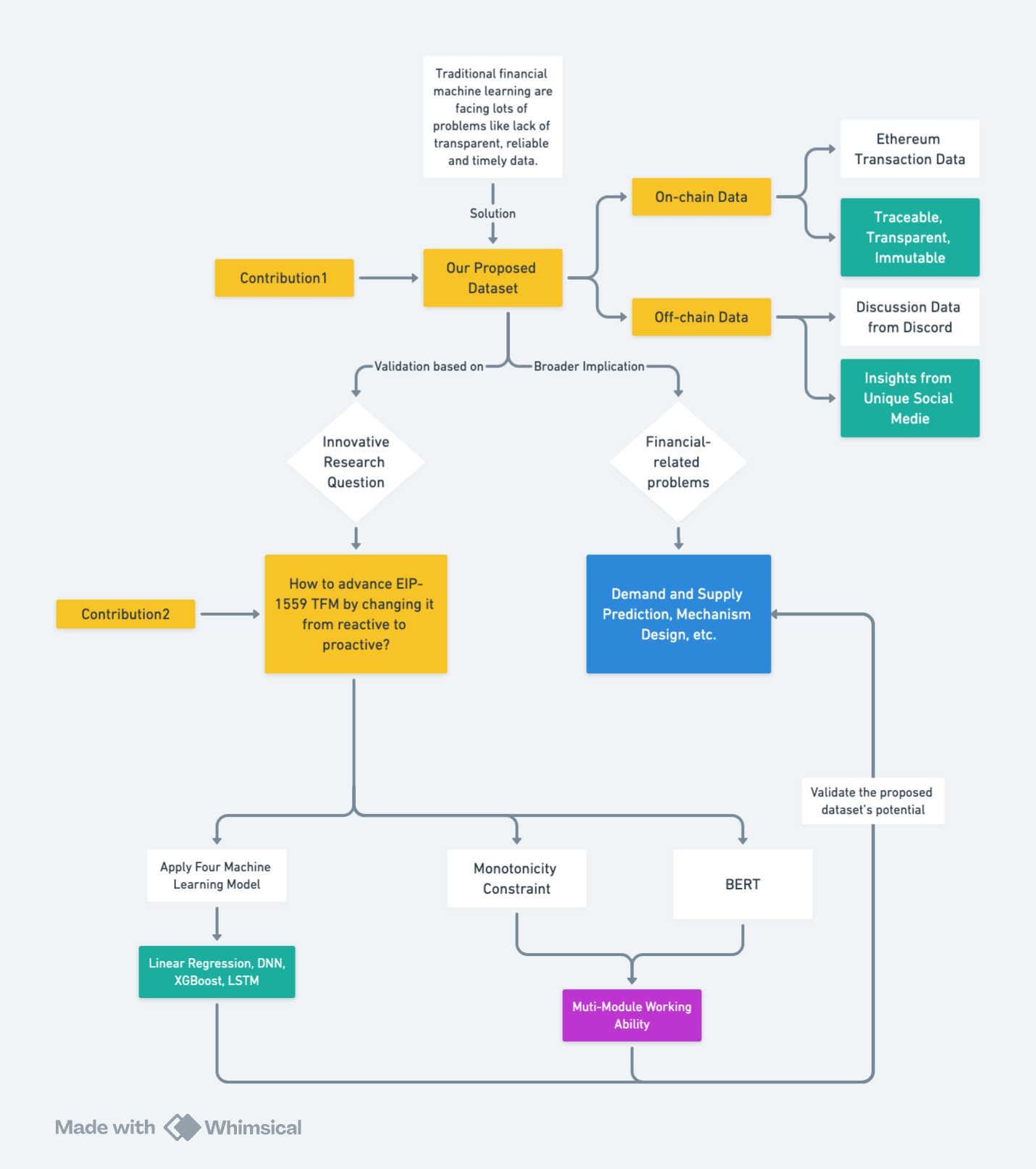} 
\caption{Accuracy at token airdrop period}
\label{Overview of the Paper}
\end{figure*}

\section{Related Work}

\subsection{The Potential of Blockchain Data}
Blockchain's diverse application scenarios provide a robust foundation for addressing challenges in financial machine learning by ensuring data integrity, security, and reliability. Its distributed, transparent, and immutable nature aligns well with the requirements of financial machine-learning models. Leading companies like Deloitte and Accenture have adopted blockchain to store and manage healthcare and medical data, demonstrating its secure data handling capabilities \cite{10.1093/jamia/ocx068}. Additionally, blockchain technology has been extended to the gaming industry, where it secures in-game assets by recording them in a distributed ledger, thereby preventing falsification \cite{10.1145/3311350.3347178}. Ma et al. proposed BlockBDM \cite{8936349}, a decentralized trust management and secure usage model for IoT big data based on public blockchain, highlighting blockchain's capability to securely handle large-scale data. Similarly, Liang et al. introduced CoopEdge, a blockchain-based decentralized platform designed to support cooperative edge computing, addressing trust issues inherent in edge computing environments \cite{10.1145/3442381.3449994}. Furthermore, the application of blockchain in the smart grid cyber layer enhances cybersecurity across various aspects, including measurement and control, data aggregation, data management, and system operation \cite{9103603}. These examples collectively underscore the security, transparency, and reliability of blockchain data, establishing it as an ideal solution for ensuring the integrity and trustworthiness of financial datasets used in machine learning models.

\subsection{Gas Price Prediction and Innovation}

Presently, there exists a plethora of scholarly endeavors dedicated to the prediction of gas prices using machine learning models.
Mars et.al.’s research has shown that the models implementing Long-Short Term Memory (LSTM) and Gated Recurrent Unit (GRU) outperform the Prophet model and the gas price oracle Geth \cite{9529485}. Butler and Crane conduct research comparing and combining different machine learning models (Direct-Recursive Hybrid LSTM, CNN-LSTM, and Attention-LSTM) to estimate the gas price in the next 5 minutes \cite{math11092212}. Chuang and Lee proposed that the Gaussian process model has a better gas price prediction when experiencing great fluctuation in transaction volumes, recommending a hybrid model consisting of both the Gaussian process and GasStation-Express \cite{deep-bdb2021}. Lan et al. propose a model combining both LSTM and XGBoost to predict the Ethereum gas price based on the data in Mempool under the EIP-1995 mechanism \cite{9973526}.

In addition, a multitude of factors are identified as exerting substantial influence on gas prices. Liu et al. utilized a regression-based gas price predicting approach (MLR) to predict the transaction fee in the next block and proposed five significant factors regarding prediction (i.e., difficulty, block gas limit, transaction gas limit, ether price, and miner reward) \cite{9045840}. Muminov et al. implemented the DeepAR model to predict Ethereum gas price and identified direct factors (e.g., seasonal variations, transaction volumes, transaction values, the number of token transactions, price, and the amount of gas used per block) as well as indirect factors (e.g., market trends, regulatory developments, and investor sentiment) \cite{muminov2024enhanced}.

These studies have predominantly focused on post-adjustment analysis, offering practical insights for guiding users' future transaction decisions. However, a significant gap exists in the research concerning the precise prediction of gas usage in subsequent blocks. As a result, there is currently no established benchmark for proactive EIP-1559-based fee adjustments. To address this gap, our research aims to develop an predictive model benchmark for gas usage in the next block. By establishing this benchmark, we create a foundation for proactive gas usage prediction, thereby facilitating future exploration and contributing to the optimization of the EIP-1559 framework. This advancement will shift the focus from post-adjustment to pre-adjustment strategies, enhancing the overall efficiency and effectiveness of dynamic transaction fee mechanisms.

\subsection{Other Blockchain-related Dataset}

\begin{table*}[ht]
\centering
\renewcommand{\arraystretch}{1.2} 
\setlength{\tabcolsep}{8pt} 

\begin{tabular}{|>{\raggedright\arraybackslash}p{2cm}|>{\raggedright\arraybackslash}p{6.5cm}|>{\raggedright\arraybackslash}p{3.5cm}|}
\hline
\textbf{Conference Abbreviation} & \textbf{Paper Title} & \textbf{Research Focus} \\
\hline
CAV & SolCMC: Solidity Compiler’s Model Checker \cite{10.1007/978-3-031-13185-1_16} & Smart Contract Verification \\
\hline
ICDCS(CCF-B) & A Graph Diffusion Scheme for Decentralized Content Search based on Personalized PageRank \cite{9951344}& Blockchain Application \\
\hline
ICDCS(CCF-B) & Distributed Runtime Verification of Metric Temporal Properties for Cross-Chain Protocols\cite{ganguly2022distributedruntimeverificationmetric} & Blockchain Analysis \\
\hline
ICDCS(CCF-B) & Monitoring Data Requests in Decentralized Data Storage Systems: A Case Study of IPFS\cite{balduf2022monitoringdatarequestsdecentralized} & Blockchain Analysis \\
\hline
INFOCOM & Blockchain Based Non-repudiable IoT Data Trading: Simpler, Faster, and Cheaper \cite{9796857} & Blockchain Performance Optimization \\
\hline
INFOCOM & BrokerChain: A Cross-Shard Blockchain Protocol for Account/Balance-based State Sharding\cite{9796859} & Blockchain Performance Optimization \\
\hline
INFOCOM & Payment Channel Networks: Single-Hop Scheduling for Throughput Maximization \cite{9796862} & Blockchain Performance Optimization \\
\hline
NSDI & DispersedLedger: High-Throughput Byzantine Consensus on Variable Bandwidth Networks\cite{276936} & Consensus Protocol \\
\hline
ICDE & Boggart: An Approximation Algorithm for Anonymity-Aware Output Decomposition in Blockchain Mixing Services \cite{9835595} & Blockchain Privacy Protection \\
\hline
ICDE & BlockOPE: Efficient Order-Preserving Encryption for Permissioned Blockchains \cite{9835603} & Blockchain Data Security \& Performance Optimization \\
\hline
ICDE & vChain+: A Searchable Blockchain System with Improved Query Performance and Public Key Management \cite{9835165}& Blockchain Query Performance Enhancement \\
\hline
Others & EX-Graph: A Dataset Authentically Linking Ethereum and Twitter \cite{wang2024exgraph} & Blockchain Data for Social Network Research  \\
\hline
Others & Chartalist: Blockchain Network Dataset for Account Relationship Analysis \cite{shamsi2024chartalist} & Blockchain Account Network \\
\hline
\end{tabular}

\caption{Summary of Blockchain-Related Research Papers}
\label{table:blockchain_research}
\end{table*}

The papers previously accepted by ICDE primarily focus on distributed ledgers and blockchains, with an emphasis on data privacy and security. For example, Ni et al. address the anonymity-aware output decomposition (AA-OD) problem by proposing an approximation algorithm, Boggart, to minimize the number of decomposed outputs while preserving transaction privacy in blockchain mixing services \cite{9835595}. Similarly, Chen et al. introduce BlockOPE, an efficient order-preserving encryption scheme for permissioned blockchains, enhancing query functionality and performance while maintaining security through parallel processing and an adaptive lightweight client cache\cite{9835603}. Wang et al. present vChain+, a searchable blockchain system that improves query performance and public key management by utilizing sliding window accumulators and tree-based indexes \cite{9835165}. However, these studies do not explicitly explore or utilize the financial attributes of blockchain-related datasets. In contrast, our research is particularly novel, as it focuses on the application of blockchain datasets in the financial sector.

There are some blockchain-based datasets being proposed to solve some financial problems. Wang et al. introduce EX-Graph, a dataset that authentically links Ethereum and Twitter\cite{wang2024exgraph}. However, the blockchain part they integrated only included OpenSea transaction data. OpenSea is just one application on Ethereum, and the NFT-related transactions it covers represent only a small portion of the overall transactions on Ethereum. Our data is sourced from Ethereum Layer 1, and the scale of our dataset is significantly larger. Shamsi et al. also tried to provide a blockchain-based dataset, but they focused on the network between accounts\cite{shamsi2024chartalist}, which is quite different from our study on supply and demand prediction and mechanism designs in financial markets. Thus, even though it seems that  similar datasets have been proposed previously, the scale and  the usage scope of our dataset are totally different and more comprehensive.

\section{Data}

As mentioned before, we propose a dataset that incorporate the on-chain data and off-chain data.

For the on-chain data, we include Ethereum's blockchain data, which includes details such as timestamps, block numbers, hashes, parent hashes, transactions, etc., which is extracted from Ethereum using Google BigQuery. For the purpose of predicting gas usage in forthcoming blocks, we retain only the pertinent features: timestamp, gas limit, gas used, and base fee. We exclude other variables such as transaction numbers, despite their high correlation with gas usage, based on our specific research focus. Furthermore, our study acknowledges the impact of token airdrops on transaction engagement levels for both recipients and non-recipients. According to Guo\cite{guo2023spillover}, token airdrops can significantly influence engagement, resulting in pronounced gas usage volatility and subsequent base fee fluctuations. Consequently, our analysis is bifurcated into two distinct periods. The first period examines the ARB token airdrop, the most substantial airdrop event in 2023, which occurred from March 21 to April 1 and comprised 78,290 blocks. The second period, devoid of significant fungible token airdrop activities, extends from June 1, 2023, to July 1, 2023, encompassing 213,244 blocks. This temporal delineation allows for a comprehensive analysis of the effects of major airdrop events on Ethereum’s gas fee dynamics.

For the off-chain data, we include users' discussion text from Discord. Discord hosts vibrant crypto discussions ranging from market analysis to technical debates, yet remains underexplored for sentiment analysis, unlike platforms like Twitter and Reddit, where extensive studies in cryptocurrency sentiment research exist (e.g., Kraaijeveld \& De Smedt, 2020 \cite{kraaijeveld2020predictive}; Mohapatra et al., 2019 \cite{mohapatra2019kryptooracle}; Khan, 2022 \cite{khan2022business}). Another reason we chose to use Discord data is its accessibility: Twitter has closed its API, limiting our ability to obtain large-scale discussion data from that platform. Moreover, Discord serves as a community hub for many crypto communities, attracting both professionals and enthusiasts interested in cryptocurrency. Consequently, the information on Discord is more targeted and relevant than discussions on other social media platforms. Specifically, we query the discussion text from the Binance, Uniswap, and Ethereum Dev communities in Discord. These communities are at the forefront of decentralized finance and blockchain development, offering a wealth of information that can reveal emerging trends, sentiments, and technical advancements. The discussion texts are queries from Discord using the DiscordChatExporter, which is an open-source tool provided on GitHub.

\section{Validation Method's Details}
\subsection{Variables}
\subsubsection{On Chain Variables}
Gas limit and gas target are two significant indicators in TFM. Specifically, the gas limit refers to the maximum amount of gas that can be consumed when executing smart contracts or transactions on each block. Gas target refers to the gas amount people want to achieve in one block. To ensure the efficiency of transactions, the gas target should equal half of the gas limit\cite{10.1145/3548606.3559341}. 

our approach is to predict the normalized gas used (denoted as y), which is calculated by the formula:
\begin{equation}
y=\frac{\text{gas used}-\text{gas target}}{\text{gas target}}
\end{equation}
This formula will shift the y within a range of [-1,1]. Or, in simple terms, this formula compares the actual gas used to the target gas limit, allowing us to assess how far off the gas usage is from the intended target. The $X$ is the variable used as features, containing $\alpha$ and $\beta$. The corresponding $\alpha$ and $\beta$ are calculated by the following formulas: 

\begin{equation}
\alpha=\frac{x_1}{x_2},
\beta=\text{base fee}
\end{equation}
where $x_1$ denotes the feature gas-used and $x_2$ denotes the feature gas-limit. $\alpha$ and $\beta$ are applied as the information of each block, and the length of previous data points used is denoted as k, which varies among the values of 1,2 and 3. Variation in $k$ aims to evaluate the performance model over different historical data lengths. 

\subsubsection{Off Chain Variables}

As mentioned in the Data section, we also incorporate an additional off-chain data source, specifically the discussion text from Discord. To analyze this data, we use a large language model to process English sentences or words, estimating the probability of each sentence being classified as positive, negative, or neutral, ensuring that the total probability sums to 1. After obtaining sentiment information, we organize the corpus sequentially and compute average sentiment scores over both hourly and daily intervals. This sentiment information is denoted as $\gamma$. We then synchronize the on-chain data with the off-chain sentiment using corresponding block data from the previous time chunk, ensuring that only preceding sentiment information is included in the training data.

\subsection{Use of Models}
\subsubsection{Neural Additive Model}
The Neural Additive Model (NAM), proposed by Agarwal et al. in 2021 \cite{agarwal2021neural}, offers a transparent framework for utilizing Deep Neural Networks (DNNs) to model individual or combined features. In this architecture, the outputs from all DNNs are aggregated at the final hidden layer, resulting in a unified model. Our research omits interactions between unrelated features, enabling the imposition of weak monotonicity constraints on each feature.

This model's relative transparency stems from the minimal interaction between variables, allowing for easy isolation of specific parts of the model to understand the relationships between the output and individual features. Consequently, even during periods of high fluctuation in gas usage, the model's specific outputs can be readily interpreted and justified.

\subsubsection{Monotonicity}
One of the primary concerns with using neural networks to predict gas usage in blockchain systems is the ‘black-box’ nature of the models. When Ethereum adopts a DNN-based approach for proactive EIP-1559, stakeholders must understand how predictions are generated to mitigate potential risks.  If users depend on overly optimistic predictions without comprehending the model's limitations, it can lead to substantial issues. Thus, ensuring the model's explainability and transparency, along with a clear understanding of the prediction procedures, is crucial.

To address these concerns, we can employ various strategies to enhance the interpretability of DNNs and incorporate well-established methods for time-series data manipulation. One commonly applied method for high-frequency time-series data, such as blockchain data, is the Exponential Moving Average (EMA) \cite{10.1007/978-3-642-41947-8_4}. EMA assigns greater weight to more recent data points, thereby providing a smoothed representation of the underlying trends in the data. The formula of EMA \cite{klinker2011exponential},\cite{ExponentialMovingAverage} is denoted as 
\begin{equation}
EMA= (C(P\_c - P\_p)) + P\_c
\end{equation} 
Where C is a constant, $P\_c$ and $P\_p$ are the current price and previous prices. Through iterative processes, the traditional EMA method often results in the forgetting of prior information, thereby forcing a high attention on the most recent data values. However, this approach can lead to a distortion of the original value distribution. A more refined solution is to address the increasing importance of data from earlier to more recent blocks, without entirely discarding historical information.

In this context, we propose a novel method for predicting gas usage in blockchain transactions, inspired by the concept of pairwise monotonicity as detailed by Chen \cite{chen2023address}. Unlike traditional methods like EMA, which emphasizes the forgetting of older information, our approach employs a monotonicity representation to attribute varying levels of importance to data over time. Monotonicity has demonstrated its interdisciplinary applicability, as evidenced by works such as Liu et al. \cite{liu2020certified} and Milani \cite{milani2016fast}, which focused on individual monotonicity for single variables. our method is inspired by Chen's work \cite{chen2023address} for introducing pairwise monotonicity in the financial domain. For instance, in credit scoring, past due amounts over a longer period should more significantly impact the scoring of new debt risk. Similarly, in blockchain transactions, older data points are less influential, whereas recent data points are more critical for prediction.

We apply monotonicity to the $\alpha$ feature, where changes in $\alpha$ for recent blocks result in greater variance in prediction compared to changes in distant previous data points. In the case of $k=2$, where the prediction uses data from the two previous blocks, the $\alpha$ values are $\alpha_1$ and $\alpha_2$ with values $(\alpha_1=a, \alpha_2=a)$. Given the higher importance assigned to $\alpha_2$, increasing or decreasing $\alpha_2$ by a certain amount $t$ compared with altering $\alpha_1$ by the same amount will lead to a higher variation of results. The mathematical equation can be denoted as
\begin{equation}
\begin{split}
\left|f(\alpha_1=a, \alpha_2=a)-f(\alpha_1=a+t, \alpha_2=a)\right| \leq\\ 
\left|f(\alpha_1=a, \alpha_2=a)-f(\alpha_1=a, \alpha_2=a+t)\right|
\end{split}
\end{equation}
The formal definition of pairwise monotonicity is modified from Chen's work \cite{chen2023address} where the individual monotonicity controls the positive sign of the addition. In other words, the model's output increases as the input variable increases. Hence, given $f$ as the model, he concludes that $f$ is weakly monotonic with respect to $x_\beta$ over $x_\gamma$ if

\begin{equation}
\begin{split}
 f\left(x_\beta, x_\gamma+c, \mathbf{x}_{\neg}\right) \leq f\left(x_\beta+c, x_\gamma, \mathbf{x}_{\neg}\right), \\
 \forall x_\beta, x_\gamma \text { s.t. } x_\beta=x_\gamma, \forall \mathbf{x}_{\neg}, \forall c \in \mathbb{R}^{+}.
 \end{split}
\end{equation}

This indicates that when given any value of $x_\beta$ and $x_\gamma$, adding a certain amount $c$, the $c$ adding on $x_\beta$ will result in a higher value. Since in our case, no individual monotonicity is imposed on the model, the changing of the output can both be positively correlated and negatively correlated with variables. Thereby, we varied the definition can propose a modified one, given $f$, the model, we conclude $f$ is weakly monotonic concerning $x_\beta$ over $x_\gamma$ if

\begin{equation}
\label{equa:mono}
\begin{split}
& \left|f\left(x_\beta, x_\gamma+c, \mathbf{x}_{\neg}\right) - f\left(x_\beta, x_\gamma, \mathbf{x}_{\neg}\right)\right| \\
& \quad \leq \left|f\left(x_\beta+c, x_\gamma, \mathbf{x}_{\neg}\right) - f\left(x_\beta, x_\gamma, \mathbf{x}_{\neg}\right)\right| \\
& \quad \forall x_\beta, x_\gamma \text { such that } x_\beta=x_\gamma, \forall \mathbf{x}_{\neg}, \forall c \in \mathbb{R}.
\end{split}
\end{equation}

Under this weak monotonicity definition, we ensure more information is addressed on the nearer data point, enhancing its transparency and explainability.

\subsubsection{FinBert Model}
The extraction of sentiment from the data is conducted using FinBert, a model proposed by Dogu and Araci in 2019 \cite{araci2019finbert}. FinBert is a BERT-based architecture specifically trained on financial data sets, including the Financial PhraseBank, TRC2-financial, and FiQA Sentiment. This training enables FinBert to achieve state-of-the-art performance in FiQA sentiment scoring. In our research, we leverage FinBert to predict sentiment in our text data, ensuring accurate sentiment analysis aligned with financial contexts.

\section{Results}

\subsection{Experiments with General Models}

To validate the applicability of our dataset for innovative financial research, we initially evaluated four machine learning algorithms—linear regression, DNN, XGBoost, and LSTM—during two distinct periods: the ARB token airdrop and a standard non-airdrop period. The ARB airdrop period, comprising 78,290 data points, allowed us to test the dataset’s robustness across diverse prediction techniques.

Each algorithm demonstrated substantial predictive accuracy, with the DNN model achieving superior performance in 23 out of 24 trials. The highest accuracy was attained when using a 10-timestep lookback, incorporating both $\alpha$ and $\beta$ features as regressors, indicating the DNN’s capacity to capture complex temporal relationships effectively.

\begin{figure}[ht]
\centering
\subfigure[Loss at token airdrop period]{
    \includegraphics[width=6.5cm]{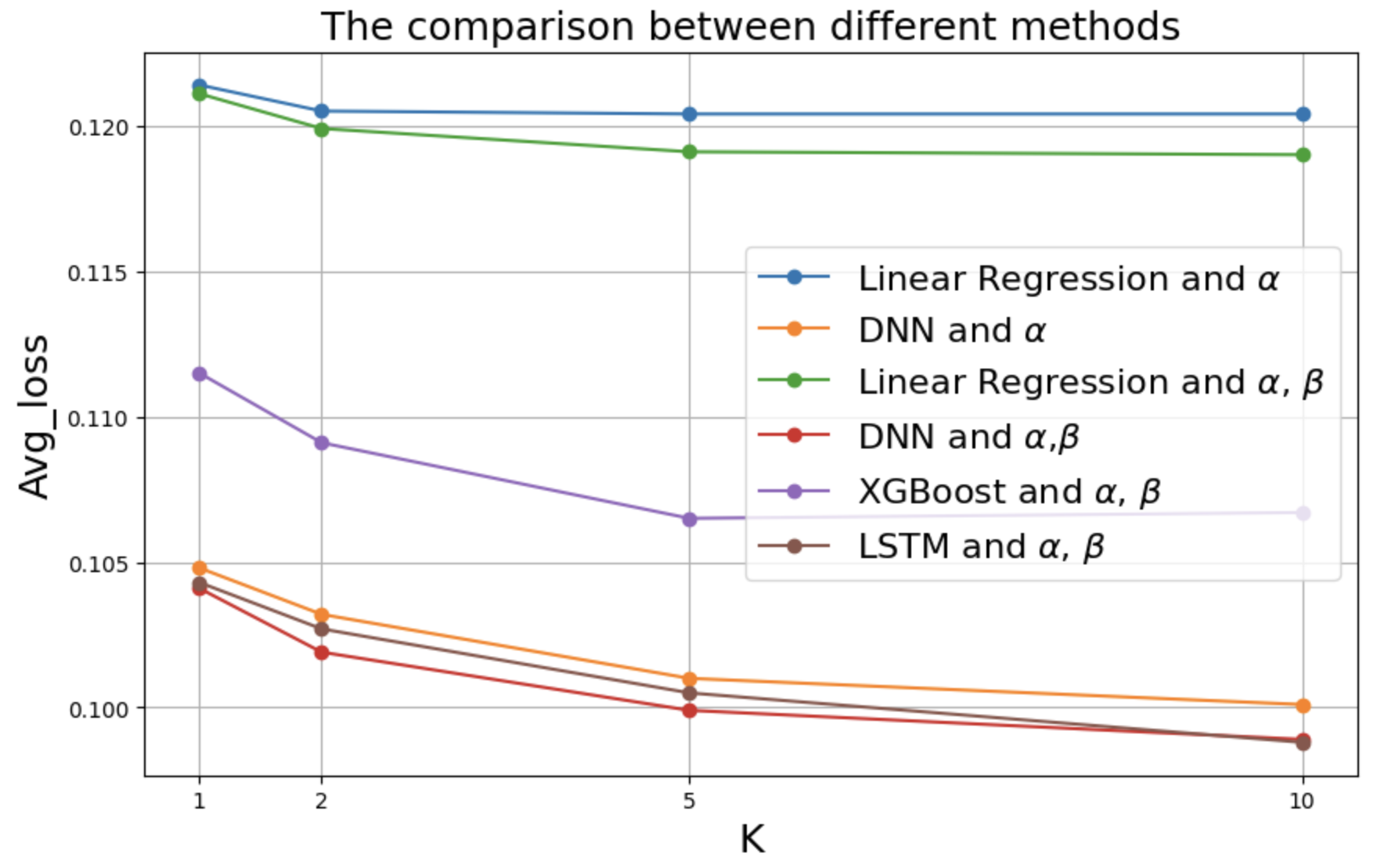}
    \label{Accuracy compare1}
}
\subfigure[Variance at token airdrop period]{
    \includegraphics[width=6.5cm]{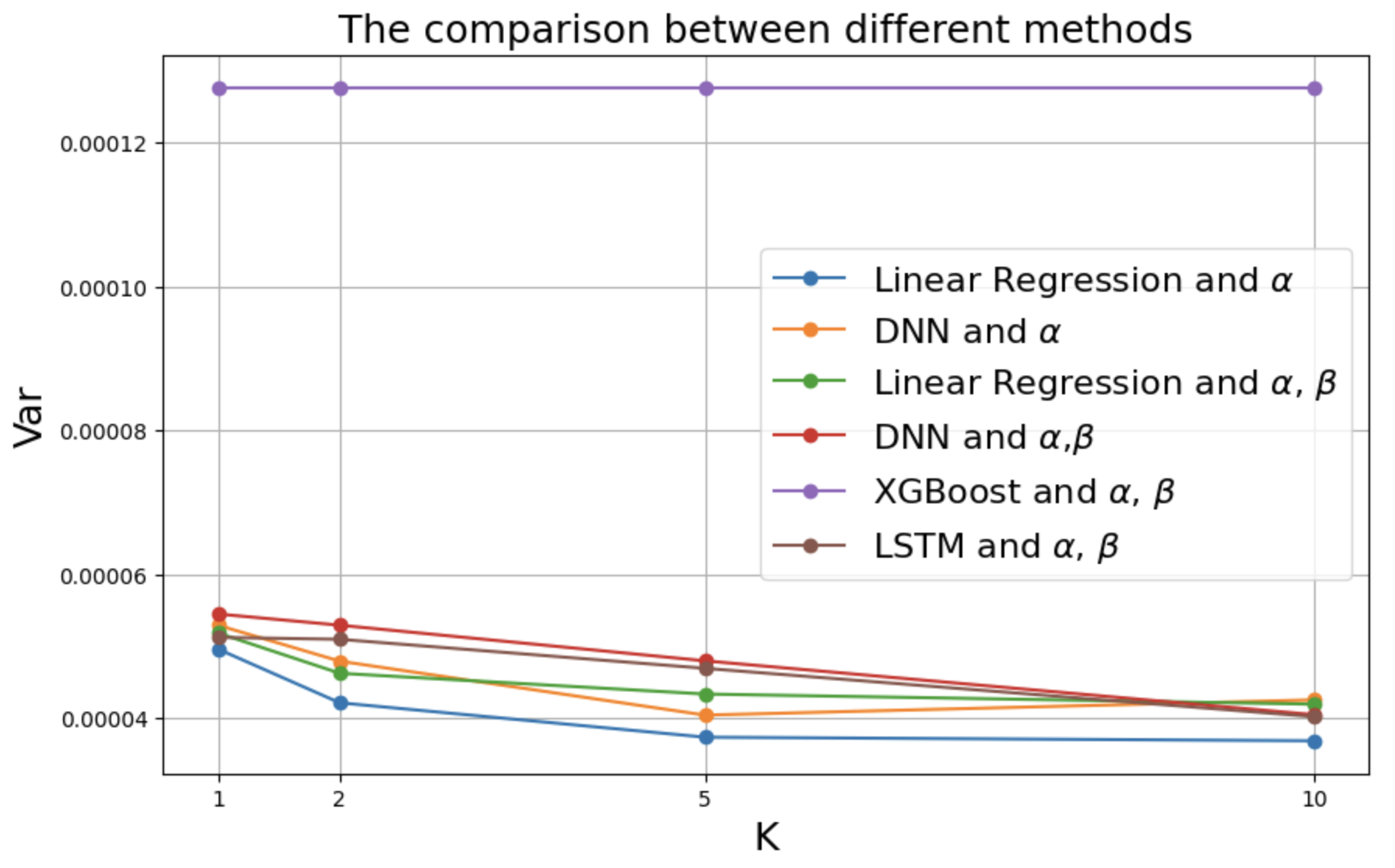}
    \label{Variance compare1}
}
\caption{Comparison of methods at Token airdrop period}
\end{figure}

A similar experimental setup was used for the normal, non-airdrop period to assess the dataset's performance in stable conditions. The predictive accuracy decreased slightly across models during this extended timeframe, with DNN again showing resilience to performance drops. While Linear Regression, XGBoost, and LSTM exhibited increased errors and variance, DNN maintained only a marginal reduction, reinforcing its suitability for handling extended prediction intervals in volatile block data.

\begin{figure}[ht]
\centering
\subfigure[Loss at normal period]{
    \includegraphics[width=6.5cm]{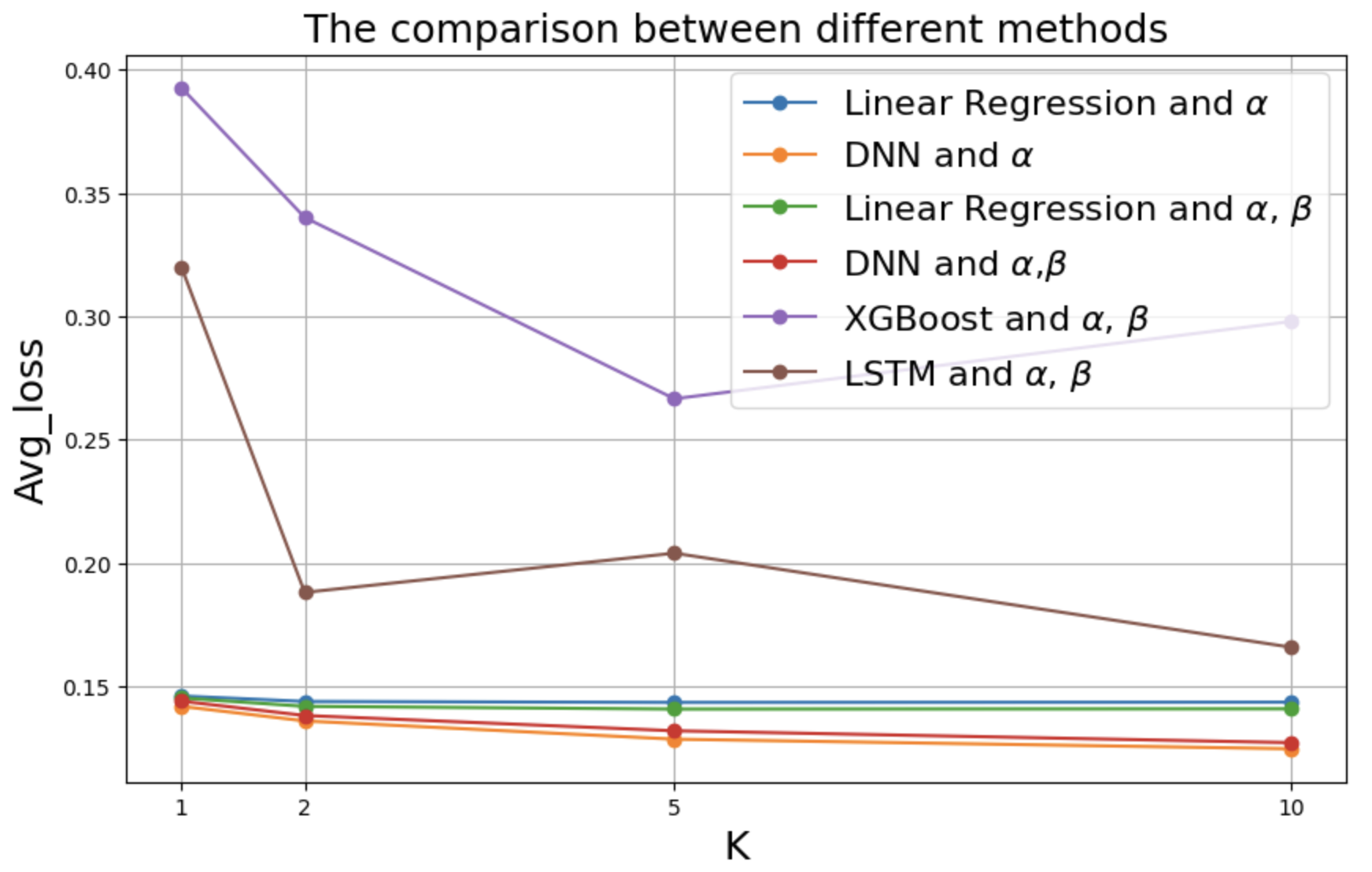}
    \label{Normal Accuracy compare}
}
\subfigure[Variance at normal period]{
    \includegraphics[width=6.5cm]{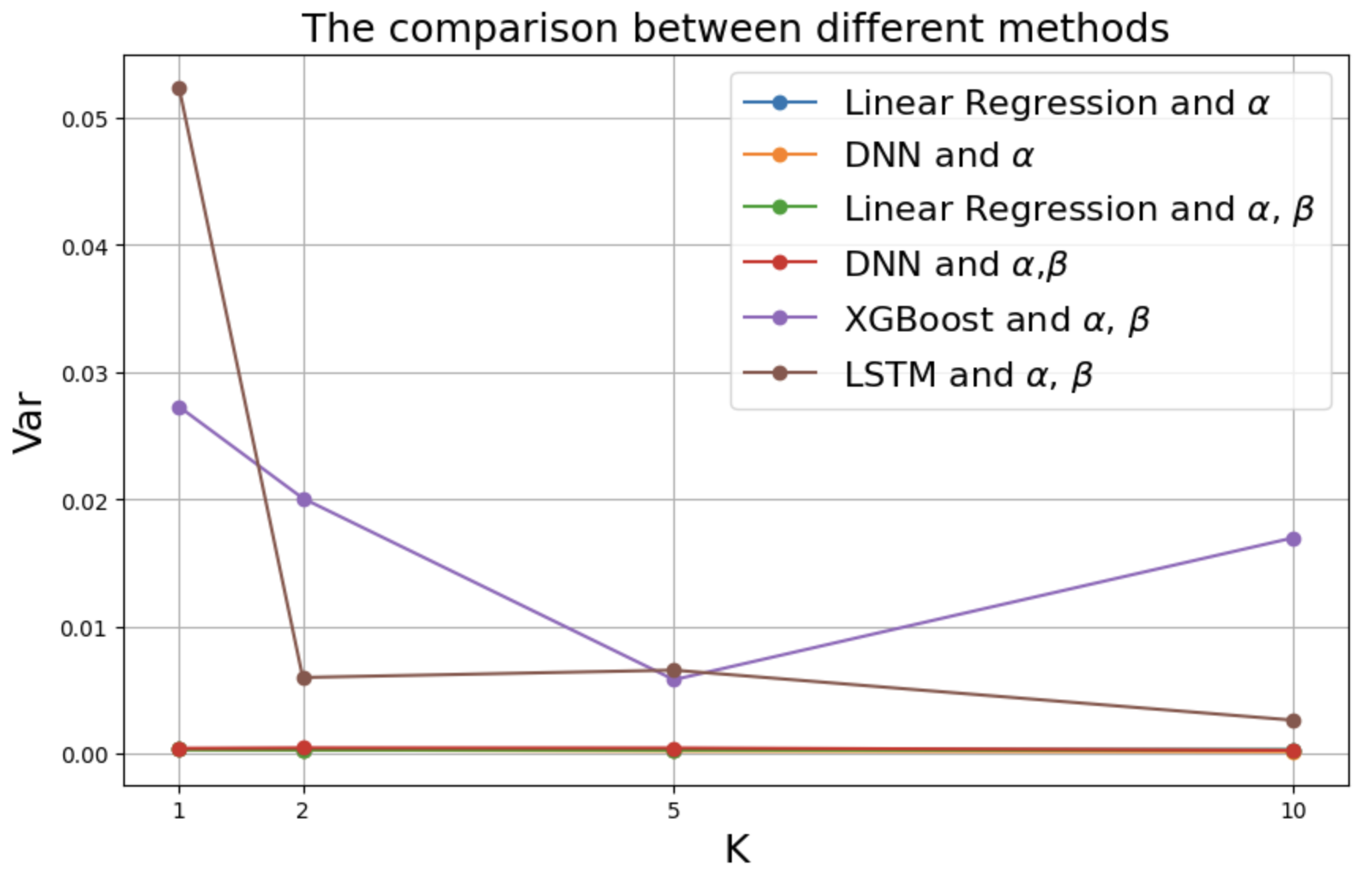}
    \label{Normal Variance compare}
}
\caption{Comparison of methods at normal period}
\end{figure}

The following Plot~\ref{Normal Accuracy compare} and Plot~\ref{Normal Variance compare} visualize each method's average mean square error and variance. The results indicate that DNN consistently performs well with low error and variance. Thereby, we continue the subsequent experiment on a structure based on the DNN structure.

The consistent performance of the DNN model across both periods led to its selection as the basis for further experiments, underscoring this dataset’s potential for supporting advanced financial prediction models under varying market conditions.

\subsection{Experiments with Monotonicity}
To demonstrate the scablity of this dataset, we employ various DNN structures, specifically the Neural Additive Model (NAM) proposed by Agarwal et al. (2021)\cite{agarwal2021neural}. We utilized the NAM model due to its inherent transparency characteristic as well as the ability to isolate variables, facilitating the imposition of monotonicity constraints on specific features.  The model is trained on data from two distinct periods, achieving weak pairwise monotonicity over the $\alpha$ feature. In the first step, standard training is conducted to enable the model to learn from the data. In the second step, we impose monotonic constraints. Specifically, discretizing the $\alpha$ feature into fractions ranging from 0 to 1 with fine intervals. Using the function described in Equation \ref{equa:mono}, we add a small value $c$ to generate a pair of inputs that must satisfy a strict inequality. The violation amount of that input, combined with the Mean Squared Error (MSE), forms a new loss function. The objective is to ensure the loss caused by the monotonicity violations becomes zero.

\begin{figure*}[htbp] 
    \centering
    \begin{minipage}{0.32\textwidth}
        \centering
        \includegraphics[width=\linewidth]{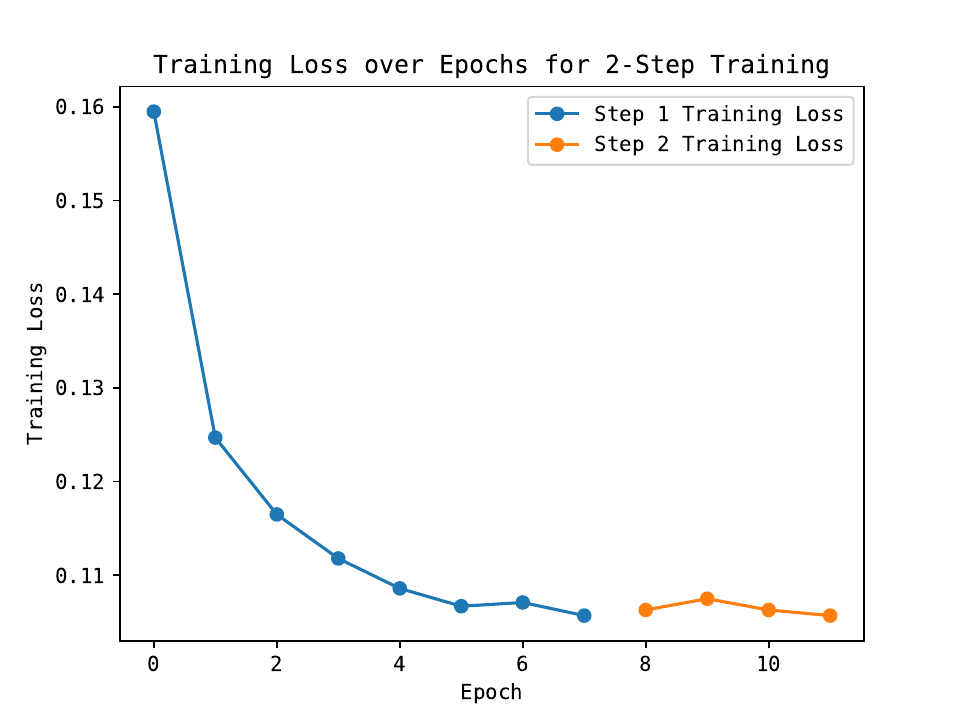}
        \caption{The two-step training loss}
        \label{fig:training loss}
    \end{minipage}\hfill
    \begin{minipage}{0.32\textwidth}
        \centering
        \includegraphics[width=\linewidth]{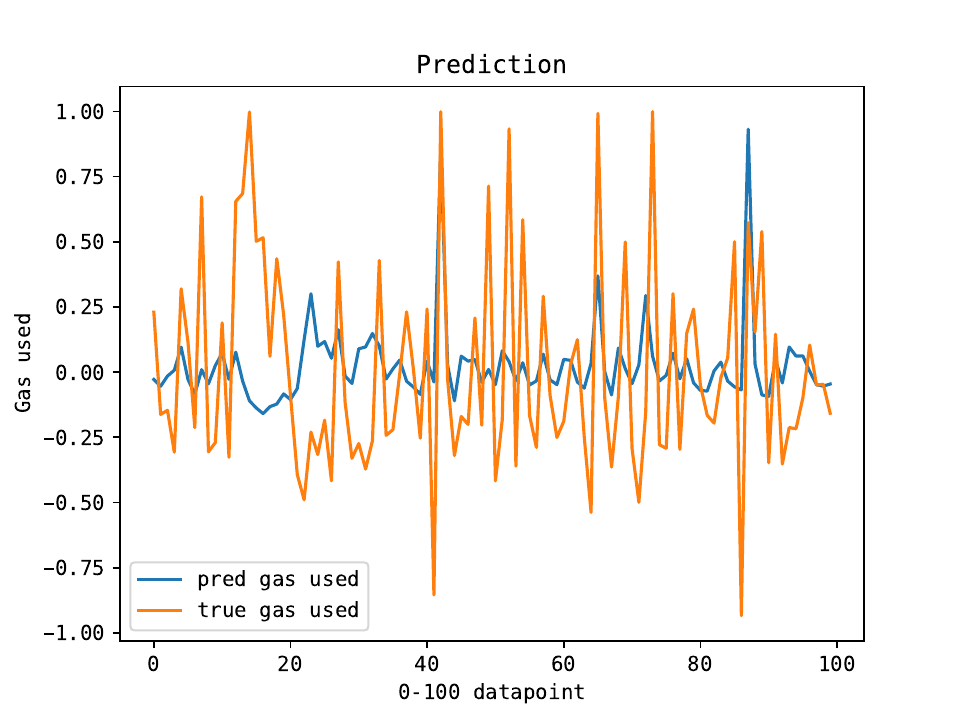}
        \caption{Prediction without monotonicity}
        \label{fig:pred_withnomono}
    \end{minipage}\hfill
    \begin{minipage}{0.32\textwidth}
        \centering
        \includegraphics[width=\linewidth]{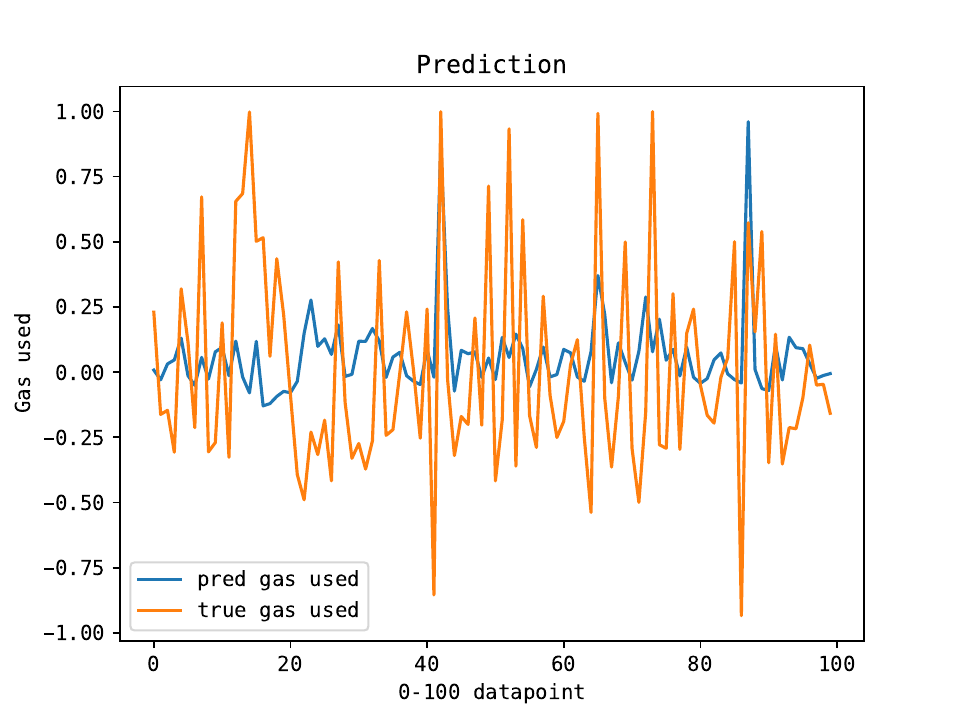}
        \caption{Prediction with monotonicity}
        \label{fig:pred_mono}
    \end{minipage}
    \caption{The NAM model with monotonicity, training and comparison}
    \label{fig:three_figures}
\end{figure*}

We only impose the monotonicity to the $\alpha$ variable since the base fee $\beta$ is generated by the Markov process from the $\alpha$. After training, we observe that with a maximum value of $k=3$, which includes three previous data points, the monotonicity constraints are satisfied without adversely affecting the loss. However, when $k=4$ or more, it becomes challenging to satisfy all monotonic constraints. Thereby, once the block number is within a constraint, we can ensure that the monotonic relationship persists and the model remains transparent. While applying monotonicity does not affect the model's loss, it enhances the model's explainability and ensures it meets the transparency requirements set by the finance director. The two-step training loss is in Plot~\ref{fig:training loss}. A comparison between the model with and without the monotonicity is shown in Figures~\ref{fig:pred_withnomono} and~\ref{fig:pred_mono}.

\subsection{Experiments with Sentiment Information}

We further explore the NAM model at k=1,2 and 3. Given the availability of both on-chain and off-chain variables, we conducted tests to determine whether the inclusion of off-chain variables, specifically sentiment analysis, enhances the model's predictability. We restricted the $k$ value to 1, 2, and 3, based on previous empirical analyses indicating that increasing $k$ beyond 3 does not significantly improve the model's performance. Additionally, since the minimum interval of the sentiment result is by an hour and the block interval is 12 seconds, indicating that the sentiment results are the same over the adjacent blocks. Hence, for all k=1,2 or 3, we only include 1 sentiment datapoint for each data. We design the experiment with 4 variable settings:

\begin{itemize}
    \item Utilizing both hour-averaged sentiment, day-averaged sentiment, and on-chain variables.
    \item Utilizing day-averaged sentiment and on-chain variables.
    \item Utilizing hour-averaged sentiment and on-chain variables.
    \item The base case, incorporates only on-chain variables.
\end{itemize}

\begin{table*}[htbp]
    \centering
    \caption{Model Performance over Two Periods}
    \label{tab:combined_table}
    \begin{tabular}{|>{\columncolor[gray]{1.0}}p{2.5cm}|p{2.5cm}|p{2.5cm}|p{2.5cm}|p{2.5cm}|}
        \hline
        \hline
        \cellcolor[gray]{0.8}
        & \cellcolor[gray]{0.8} +OC,+DS,+HS & \cellcolor[gray]{0.8} +OC,+DS,-HS & \cellcolor[gray]{0.8} +OC,-DS,+HS & \cellcolor[gray]{0.8} +OC,-DS,-HS \\
        \hline
        \multicolumn{5}{|c|}{\cellcolor[gray]{0.8}\textbf{Period 1: 03/21/2023 - 04/01/2023 (ARB-airdrop)}} \\
        \hline
        3 Timesteps & 0.10022 & 0.10150 & 0.10164 & 0.10201 \\
        \hline
        2 Timesteps & 0.10056 & 0.10249 & 0.10213 & 0.10265 \\
        \hline
        1 Timestep  & 0.10169 & 0.10190 & 0.10204 & 0.10290 \\
        \hline
        \multicolumn{5}{|c|}{\cellcolor[gray]{0.8}\textbf{Period 2: 06/01/2023 - 07/01/2023 (Normal)}} \\
        \hline
        3 Timesteps & 0.13341 & 0.15657 & 0.16142 & 0.16089 \\
        \hline
        2 Timesteps & 0.13477 & 0.15381 & 0.15806 & 0.16456 \\
        \hline
        1 Timestep  & 0.13593 & 0.15321 & 0.15459 & 0.18428 \\
        \hline
        \hline
    \end{tabular}
\caption{The notation "OC" refers to On-chain variables, while "HS" and "DS" denote Hourly Averaged Sentiment and Daily Averaged Sentiment, respectively. ‘+’ symbol indicates the inclusion of a variable in the model, whereas ‘-’ symbol denotes its exclusion. The numerical values represent the mean square error (MSE) of the model on the test dataset.}
\end{table*}

\begin{figure*}[h]
\centering
\subfigure[40 datapoints]{
    \includegraphics[width=6.5cm]{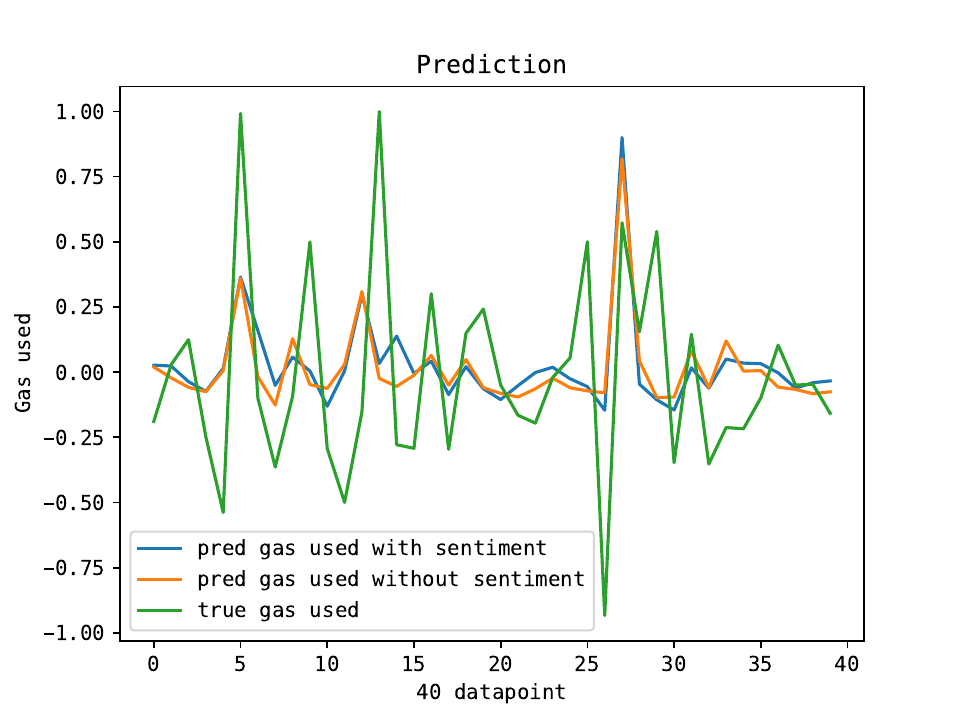}
    \label{Prediction on 40 points}
}
\subfigure[100 datapoints]{
    \includegraphics[width=6.5cm]{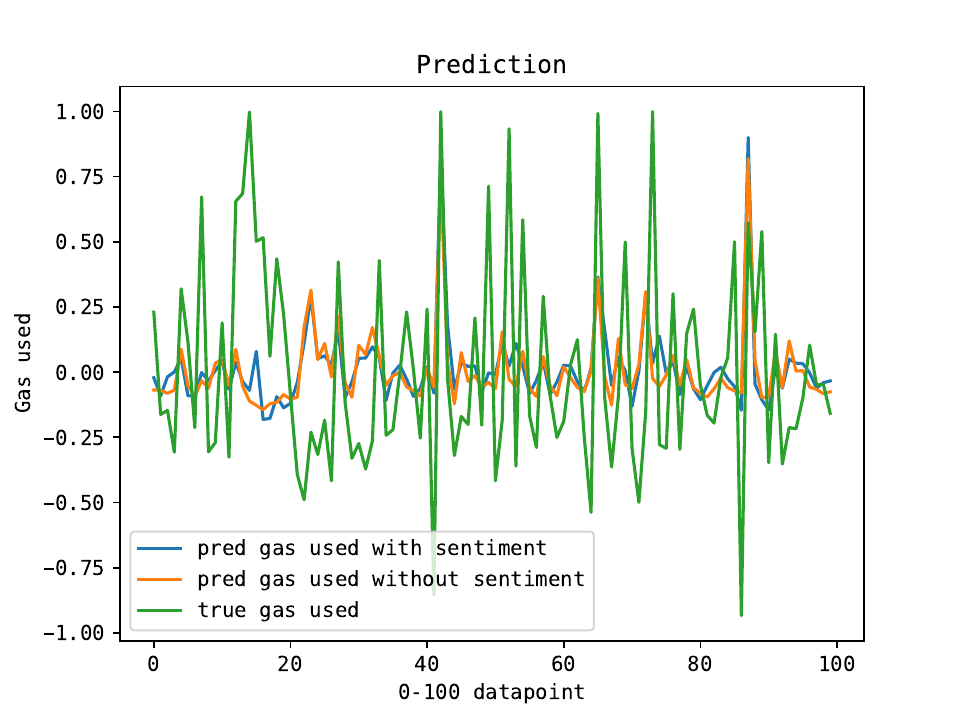}
    \label{Prediction on 100 points}
}
\caption{Comparitive prediction results of model with and without sentiments}
\end{figure*}

The prediction on the test set with 40 and 100 data points is demonstrated in Figure~\ref{Prediction on 40 points} and~\ref{Prediction on 100 points}. The sentiment analysis can marginally improve the accuracy of the model. Additionally, Adding timesteps enhances predictability when only on-chain data is used. However, when sentiment data is included, the number of needed timesteps decreases.

The results, summarized in Table~\ref{tab:combined_table}, show that sentiment addition had a minor effect on overall predictive accuracy, with slight improvements during high-activity periods, such as the ARB airdrop. The dataset’s design enabled analysis across different configurations, demonstrating flexibility in supporting diverse data types and hybrid prediction models. Nonetheless, sentiment data in its averaged form displayed limited impact, likely due to unrelated information within the corpus that reduced its relevance to gas usage patterns. However, this can be addressed by further refining the filtering methods for text data.

In summary, our findings validate the versatility and robustness of this dataset for implementing diverse machine learning techniques and financial prediction models. The experiments substantiate that the dataset not only supports various prediction methods but is also equipped to handle complex, hybrid data types, reinforcing its suitability for exploring novel financial models and enhancing blockchain transaction predictability.

\section{Conclusion and Future Study}
In conclusion, we present a novel dataset that integrates both on-chain and off-chain data, specifically designed to address key financial challenges, such as mechanism design, through advanced machine learning techniques. This dataset addresses the critical need for robust, data-driven decision-making in both blockchain and financial systems. One of the primary research questions we explore using this dataset is how to accurately predict gas usage in upcoming blocks and dynamically adjust transaction fees based on specific financial objectives, such as cost optimization or efficiency improvements. To evaluate the adaptability and effectiveness of this dataset, we applied four widely-used machine learning algorithms—Linear Regression, LSTM, Deep Neural Networks (DNNs), and XGBoost—demonstrating its flexibility for various analytical approaches. Additionally, we incorporated monotonicity and sentiment analysis to validate the dataset's ability to support multi-modular research, crucial for understanding both quantitative patterns and qualitative insights in financial markets.

Our research underscores the limitations of current blockchain data usage frameworks, particularly in their adaptability to financial applications, and highlights the need for systems capable of seamlessly integrating diverse data sources while supporting sophisticated machine learning techniques. Looking forward, this dataset sets a new direction for future research in blockchain-based financial data management. This dataset demonstrates its potential of leveraging machine learning for predictive analytics in blockchain systems, offering a foundation for enhancing transaction mechanisms and financial modeling in finance, especially in decentralized finance (DeFi). By making this dataset openly available, we aim to support the broader research community and drive further exploration into the intersection of blockchain data and machine learning to foster innovation in the financial sector.

\section*{Acknowledgment}

Luyao Zhang is supported by the National Science Foundation of China (NSFC) for the project titled “Trust Mechanism Design on Blockchain: An Interdisciplinary Approach of Game Theory, Reinforcement Learning, and Human-AI Interactions” (Grant No. 12201266). Jingfeng Chen and Wanlin Deng are supported by the Office of Academic Services and the DKU Summer Research Scholar Program, under the supervision of Prof. Dangxing Chen and Prof. Luyao Zhang, respectively.

\bibliographystyle{IEEEtran}
\bibliography{reference}
\vspace{12pt}

\end{document}